\documentstyle[epsf]{article}
\begin{document}

\renewcommand{\baselinestretch}{1.5}
\large

\begin{center}
{\Large {\bf
A Method to Calculate a Delta Function of the \\ Hamiltonian 
  by the Suzuki-Trotter Decomposition
} }
\vskip 0.2in
   T. MUNEHISA and Y.MUNEHISA
\vskip 0.2in
Faculty of Engineering, Yamanashi University

Takeda, Kofu, Yamanashi 400-8511, Japan
\end{center}
\vskip 0.3in
{\bf ABSTRACT}

 We propose a new method to calculate expectation values of
a delta function of the Hamiltonian,
$ \langle  \Psi \mid \delta( \hat{H} - E)\mid \Psi \rangle $.
Since the delta function can be replaced with a Gaussian function, 
we evaluate
 $$ \langle  \Psi \mid \sqrt{\frac{\beta}{\pi}} 
e^{ -\beta (\hat{H} - E)^2} \mid \Psi \rangle $$
with large $\beta$ adopting the Suzuki-Trotter decomposition.
Errors of the approximate calculations with the finite Trotter number
$N_t$ are estimated to be $ O(1/N_t^K)$ for the $K$th-order decomposition.
 The distinct advantage of this method is that the convergence is guaranteed
 even when the state $\mid \Psi \rangle$ contains the eigenstates whose 
energies spread over the wide range.
 
In this paper we give a full description of our method within the quantum 
mechanical physics and present the numerical results for the harmonic 
oscillator problems in one- and three-dimensional space.


\eject
\normalsize
\section{Introduction}

In the quantum physics some dynamical  quantities in the matter\cite{iitaka},
\cite{rv}
 can be expressed by
\begin{eqnarray}
% a(E) &=& {\rm Im} \{ 
 a(E) = {\rm Im} \{ 
\langle  \Psi \mid \hat{A}^\dagger \frac{1}{ \hat{H} -E 
-i \epsilon } \hat{A} \mid \Psi \rangle \} 
%\nonumber \\
% &=&  \pi 
 =  \pi 
\langle  \Psi \mid \hat{A}^\dagger \delta( \hat{H} -E)
\hat{A} \mid \Psi \rangle . 
\label{eq:Im}
\end{eqnarray}
Examples of them
are the density  of states and the forward scattering amplitude.
There are many attempts to calculate the delta function by the polynomial 
expansions such as the Chebyshev polynomial expansion\cite{delta}.
In these expansions it is assumed that the eigenvalues are situated inside 
the fixed ranges. Whether one can obtain a good convergence or not with the 
expansion, therefore, entirely depends on the state $\mid \Psi \rangle$.

In this paper we investigate a new method to numerically calculate 
the expectation value of the delta function. Our starting point is that 
the delta function can be represented by the Gaussian function
\footnote{ The Gaussian function has been used for the filter function,
where the polynomial expansion and other methods are employed to
calculate it\cite{filter}. },
\begin{eqnarray}
\langle  \Psi \mid \hat{A}^\dagger \delta( \hat{H} -E)
\hat{A} \mid \Psi \rangle
= \lim_{\beta  \rightarrow \infty }  \ 
\langle  \Psi \mid \hat{A}^\dagger \sqrt{\frac{\beta}{\pi}}
e^{-\beta( \hat{H} -E)^2  }\hat{A} \mid \Psi \rangle .
\label{eq:G}
\end{eqnarray}
It is justified in the evaluation to use 
$\sqrt {\beta / \pi} \cdot e^{ - \beta( \hat{H} -E)^2 }$ 
with large but {\em finite} $\beta$ instead of the delta function 
because $\beta ^{-1}$ can be interpreted as the resolution in the
actual observations.  
We then employ the Suzuki-Trotter decomposition
to calculate the expectation value of the Gaussian function. 
This decomposition is quite stable 
as is well known in the studies of the quantum Monte Carlo method\cite{MC}.
Extensive work has also been done to calculate $ e^{-i  \hat{H} t } $ 
by this composition\cite{natori}.

In the next section we explain our method in detail, applying it to  
the one-dimensional quantum mechanics. 
In Sec.~3 the higher-order decompositions\cite{suzuki} are described and 
Sec.~4 is devoted to estimating errors in the method.
We then present numerical results for the one-dimensional harmonic 
oscillator in Sec.~5, where the stability of the method is also demonstrated.
Sec.~6 contains applications to the three-dimensional problems
and the final section is for the summary.

\section{Methods}

Our study here will be limited to the quantum mechanics, i.e. 
to the one particle problems.
Here we describe the method in the one-dimensional case.
The Hamiltonian is given by
\begin{eqnarray}
 \hat{H} = -\frac{d^2 }{dx^2}+ V(x),
\label{eq:Hm}
\end{eqnarray}
where we adopt units of  $\hbar=1$ and $m=1/2$.
From Eq.(\ref{eq:Hm}) it follows
\begin{eqnarray}
 \hat{H}^2=\frac{d^4 }{dx^4}-\frac{d^2 }{dx^2} V(x)-V(x)\frac{d^2 }{dx^2}
+V(x)^2 .
\end{eqnarray}
Next we employ the discrete space representation,
\begin{eqnarray}
x_i = (i-1)\Delta + x_{min} \ \ \ \ (i=1, \cdots ,L), \ \ \ \ 
  \Delta = (x_{max} - x_{min})/L .
\label{eq:Sr}
\end{eqnarray}
The wavefunction $\phi(x)=\langle x \mid \phi \rangle$
 is replaced by $\phi(x_i)$, which is denoted as $\phi_i$ hereafter.
Then the differentials become 
\begin{eqnarray}
 \frac{d^2\phi(x)}{ dx^2 } \rightarrow 
\frac{1}{\Delta^2} ( \phi_{i+1}+\phi_{i-1}
-2\phi_i ),
\end{eqnarray}
\begin{eqnarray}
\frac{d^4\phi(x)}{ dx^4 } \rightarrow 
\frac{1}{\Delta^4} ( \phi_{i+2}+\phi_{i-2}
-4 \phi_{i+1}-4\phi_{i-1} 
+6\phi_i) ,
\end{eqnarray}
\begin{eqnarray}
& &V(x) \frac{d^2\phi(x)}{ dx^2 }+ \frac{d^2V(x) \phi(x)}{ dx^2 }  \rightarrow
\nonumber \\
& & \frac{1}{\Delta^2} \{[V(x_{i+1})+V(x_{i}) ] \phi_{i+1}+
 [ \ V(x_{i})+V(x_{i-1}) ]\phi_{i-1} -4V(x_i) \phi_i \},
\label{eq:D3}
\end{eqnarray}
respectively, and
we obtain a matrix $H_2$ to represent $(\hat{H} -E)^2$.
In order to calculate $e^{-\beta H_2}$ using the Suzuki-Trotter 
decomposition,
we divide $H_2$ into four matrices, 
 $ H_2^{(n)}$ $( n=1,2,3,4 )$, which are formed by aligning
the $4\times 4$ matrices along the diagonal line (See Fig.~1),
\begin{eqnarray}
H_2 = H_2^{(1)}+  H_2^{(2)}+  H_2^{(3)} + H_2^{(4)}. 
\end{eqnarray}
In the second-order decomposition 
 we approximate  $e^{-\beta H_2}$ by
\begin{eqnarray} 
%& &\{ e^{-\frac{\beta}{2N_t}H_2^{(1)} }
% e^{-\frac{\beta}{2N_t}H_2^{(2)} }
% e^{-\frac{\beta}{2N_t}H_2^{(3)} }
% \nonumber  \\
%& & e^{-\frac{\beta}{N_t}H_2^{(4)} } 
% e^{-\frac{\beta}{2N_t}H_2^{(3)} } 
% e^{-\frac{\beta}{2N_t}H_2^{(2)} }
%  e^{-\frac{\beta}{2N_t}H_2^{(1)} } \}^{N_t}
 & & [ exp(-\frac{\beta}{2N_t}H_2^{(1)} )
 exp(-\frac{\beta}{2N_t}H_2^{(2)} )
 exp(-\frac{\beta}{2N_t}H_2^{(3)} )
 \nonumber  \\
& & exp(-\frac{\beta}{N_t}H_2^{(4)} ) 
 exp(-\frac{\beta}{2N_t}H_2^{(3)} ) 
 exp(-\frac{\beta}{2N_t}H_2^{(2)} )
  exp(-\frac{\beta}{2N_t}H_2^{(1)} ) ] ^{N_t}
\end{eqnarray}
with the finite Trotter number $N_t$.
The error in this approximation is $O(1 / N_t^2)$, as is well known.

\section{Higher-order decompositions}

In Ref.\cite{suzuki} the higher-order decompositions have been discussed.
This section is to briefly refer, for later use in our calculations, 
to the fourth- and the sixth-order decompositions together with the 
second-order one.
For simplicity we describe here only the case where the operator $\hat C$ is 
the sum of the two operators $\hat{A}$ and $\hat{B}$.

The $K$th-order formula is defined by
  \begin{eqnarray}
e^{x \hat C } = 
  e^{x(\hat{A}+\hat{B})} = [ \hat{S}_K(x/N_t) ]^{N_t} +O(x^{K+1}/N_t^K).
\end{eqnarray}
For the second-order decomposition ($K = 2$) we have
\begin{eqnarray}
\hat{S}_2(x) = e^{\frac{x}{2}\hat{A} }e^{x \hat{B} }e^{\frac{x}{2}\hat{A}}.
\label{eq:S2}
\end{eqnarray}
Using $\hat S_2$ the fourth-order decomposition ($K = 4$) is given by 
\begin{eqnarray}
\hat{S}_4(x) = \hat{S}_2(p_2x)^2 \hat{S}_2((1- 4 p_2)x) \hat{S}_2(p_2x)^2, 
\label{eq:S4}
\end{eqnarray}
with $p_2=(4-4^{1/3})^{-1} = 0.4144907717943757$, 
while the sixth-order one ($K = 6$) is represented as
\begin{eqnarray}
\hat{S}_6(x) = \hat{S}_2(p_1x)^2 \hat{S}_2(p_2x)^2 \hat{S}_2(p_3x)^2 
\hat{S}_2(p_4x)^2 \hat{S}_2(p_3x)^2 \hat{S}_2(p_2x)^2 \hat{S}_2(p_1x)^2 
\label{eq:S6}
\end{eqnarray}
with $p_1 = 0.3922568052387732$, $p_2 = 0.1177866066796810$, \\ 
$p_3 = -0.5888399920894384$ and $p_4= 0.6575931603419684$.

Since we have already obtained $\hat{S}_2$ for our problem in the previous 
section, it is straightforward to apply these higher-order decompositions 
to our calculations. 

\section{Errors in the Suzuki-Trotter decomposition}

In this section we roughly estimate errors in our method using eigenstates
 $ \mid \psi_i \rangle $ and eigenvalues $\lambda_i$
for the hermitian operator $\hat{C}$, namely
\begin{eqnarray}
 \hat{C}\mid \psi_i \rangle = \mid \psi_i \rangle \lambda_i ,
\end{eqnarray}
with $\langle \psi_i \mid \psi_j \rangle = \delta_{ij}$.

For the operator $\hat{S}_K(\delta)$
in the $K$th-order formula, where $\delta$ denotes some small $c$-number,
we have 
$$ \hat{S}_K(\delta)= e^{\delta\hat{C}} + \hat{E}_K , $$
where $\hat{E}_K$ is $O(\delta^{K+1})$.
Then we apply the decomposition for $ e^{\delta N_t \hat{C}}$ with a fixed   
$\delta N_t$, 
\begin{eqnarray}
[\hat{S}_K(\delta)]^{N_t} &=& [ (e^{\delta\hat{C}} + \hat{E}_K )]^{N_t}
\nonumber \\
&=& e^{N_t\delta\hat{C}} + \sum_{k=0}^{N_t-1}e^{ k\delta\hat{C}}
\hat{E}_Ke^{ (N_t-k-1)\delta\hat{C}}
+O(\delta^{2(K+1)})  .
\label{eq:fy}
\end{eqnarray}
Note that the $\delta$ should be $O(1/N_t)$ here.
In order to go on with the estimation we neglect the operators of 
$O(\delta^{2(K+1)})$ in Eq.(\ref{eq:fy}), although this is not a rigorous 
procedure.
With this approximation we obtain 
\begin{eqnarray}
\langle \psi_j \mid [ \hat{S}_K(\delta)]^{N_t} \mid \psi_i \rangle 
& \simeq & \langle \psi_j \mid 
 e^{N_t\delta\hat{C}} + \sum_{k=0}^{N_t-1}e^{ k\delta\hat{C}}
\hat{E}_Ke^{ (N_t-k-1)\delta\hat{C}}  \mid \psi_i \rangle
\nonumber \\
&=& \langle \psi_j \mid  e^{ N_t\delta \lambda_i} + 
 \sum_{k=0}^{N_t-1}e^{ k\delta\lambda_j}
\hat{E}_K e^{ (N_t-k-1)\delta\lambda_i}  \mid \psi_i \rangle  
\nonumber \\
&=& e^{ N_t\delta \lambda_i}\delta_{ij} +
 \sum_{k=0}^{N_t-1}e^{ k\delta\lambda_j }e^{ (N_t-k-1)\delta\lambda_i} 
\langle \psi_j \mid  \hat{E}_K\mid \psi_i \rangle .
\label{eq:er1}
\end{eqnarray}
For $j=i$ it reads
\begin{eqnarray}
  \langle \psi_i \mid [ \hat{S}_K(\delta)]^{N_t} \mid \psi_i \rangle
 \simeq e^{ N_t\delta \lambda_i }  [ 1  +   \langle 
 \psi_i \mid  \hat{E}_K\mid \psi_i \rangle  N_t e^{ -\delta\lambda_i}].
\label{eq:ecase}
\end{eqnarray}
When $j \ne i$, on the other hand, we have  
\begin{eqnarray}
 \langle \psi_j \mid [ \hat{S}_K(\delta)]^{N_t} \mid \psi_i \rangle &\simeq&
 \sum_{k=0}^{N_t-1}e^{ k\delta\lambda_j }e^{ (N_t-k-1)\delta\lambda_i} 
\langle \psi_j \mid  \hat{E}_K\mid \psi_i \rangle  
\nonumber \\
&=& \langle \psi_j \mid  \hat{E}_K\mid \psi_i \rangle
\frac{ e^{ N_t\delta\lambda_i} - e^{ N_t\delta\lambda_j}  }
{e^{ \delta\lambda_i} - e^{ \delta\lambda_j} } .
\label{eq:necase}
\end{eqnarray}
Eq.(\ref{eq:necase}) suggests that the error is
$\sim \langle \psi_j \mid  \hat{E}_K\mid \psi_i \rangle 
N_t e^{(N_t -1)\delta \lambda_i} \sim \delta^K$ 
if $ \delta(\lambda_j - \lambda_i) \approx 0$, while for the case 
$\lambda_j \gg \lambda_i $ it is estimated by
$\langle \psi_j \mid  \hat{E}_K\mid \psi_i \rangle
e^{ (N_t-1)\delta\lambda_j} \sim \delta^{K+1}$.

\section{Results in one dimension}

Now we present our results for the one-dimensional harmonic 
oscillator whose potential is
\begin{eqnarray}
 V(x) = \lambda  x^2 ,
\end{eqnarray}
where we fix $\lambda=1/4$ so that the energy eigenvalues are
$(k-1/2) \ \ ( k=1,2,3,\cdots )$ in the continuum limit.
In our calculations with the discrete space
representation using Eqs.~(\ref{eq:Sr}) $\sim$ (\ref{eq:D3}) 
we scale the matrix $H_2$ by $\Delta^4$, which demands that 
the energy should be $\Delta^2 (k-1/2) $ instead of $(k-1/2)$.
Throughout this section 
we set $x_{min} = -25$, $ x_{max} =  25$ and $L=500$ in Eq.~(\ref{eq:Sr}).

Let $\mid \phi^{(k)} \rangle$ be the eigenstate of the Hamiltonian 
with the eigenvalue $E_k$, for which the ordering $E_k < E_{k+1} $ is assumed.
First, following to Eq.(\ref{eq:er1}) and employing the fourth-order 
decomposition, we estimate the errors of 
$ \langle \phi^{(k)} \mid e^{-\beta(\hat{H}-E)^2 } \mid \phi^{(20)} \rangle$
with $\beta= 100$ for several states $\mid \phi^{(k)} \rangle$.
In order to estimate errors owing to the decompositions only,
we compare our numerical results with those obtained by the  
exact diagonalization of the matrix $H_2$.
In Fig.~2 we plot the results as a function of the Trotter number $N_t$, 
which clearly indicate that the errors decrease as $O(N_t^{-4})$.
Fig.~3 is to compare the errors estimated with $N_t=1000$ and those with 
$N_t=2000$ for each $k$ up to 40.
Here we see that we can make the errors quite small 
by increasing $N_t$.  Note that the largest error is found 
at $k=20$, which is the case described by the Eq.~(\ref{eq:ecase}).

Next we calculate 
\begin{eqnarray}
 I(E) \equiv \langle \phi \mid e^{-\beta (\hat{H}-E)^2} \mid \phi \rangle ,
\end{eqnarray}
using the wavefunction $\phi_i$ which is parametrized by
\begin{eqnarray}
\phi_i = \phi_i^{(1)} C_1 + ... + \phi_i^{(L)} C_{L}, \ \ \ \ \
C_1^2 +...+C_{L}^2 =1,
% & & \phi_i = \phi_i^{(1)} C_1 + ... + \phi_i^{(L)} C_{L}, \\
% & &C_1^2 +...+C_{L}^2 =1, \nonumber
\end{eqnarray}
with the discretized wavefunctions $ \phi^{(k)}_i \equiv 
\mid \phi^{(k)} (x_i) \rangle$.
Here we use two wavefunctions as $\phi_i$.
One of them, which we will call the wavefunction (A), is given by
$$ C_k=\frac{1}{\sqrt{20}} \ \ for  \ \ k \leq 20, \ \ \ \ \
C_k=0 \ \ for  \ \  k \geq 21. $$
For the wavefunction (B), which is useful to study the effects due to 
the highly excited states, we set
$$ C_k=C(\frac{1}{\sqrt{20}}+\frac{1}{100}) \ \ for  \ \ k \leq 20, $$
$$ C_k=C/100 \ \ for  \ \ 21 \leq k \leq 100, \ \ \ \ \
 C_k=0 \ \ for  \ \  k \geq 101, $$
where $C$ is the normalization factor. 
In Fig.~4 we plot the error ratios for the wavefunction (A) 
with the second-, the fourth- and the sixth-order decompositions as a 
function of the effective Trotter number $N'_t$, which is the Trotter 
number multiplied by the number of $\hat S_2$'s in Eqs.~(\ref{eq:S2}),
~(\ref{eq:S4}) and (\ref{eq:S6})\footnote{Namely $N'_t = N_t$ 
$(N'_t =5N_t$, $N'_t =14N_t$)
for the second-order (fourth-order, sixth-order) decomposition.}.
Results for the wavefunction (B) with the fourth-order decomposition 
are also plotted for a comparison. 
We see that the errors from each decomposition decrease as $O(1/N_t^K)$ 
in agreement with the theoretical expectation. 
It should be noted, however, that the fourth-order decomposition is more 
recommendable than the sixth-order one because the error from the
latter turns out to start decreasing only for the large value of $N_t$.

Fig.~5 is to demonstrate that the operator $e^{-\beta(\hat{H}-E)^2}$ can 
indeed pick up the states with the energy $\sim E$. We calculate
$ L \cdot \langle \phi^{(k)} \mid e^{-\beta (\hat{H}-E)^2} \mid \phi \rangle$
$(1 \le k \le L)$ 
using the wavefunction $\phi_i \equiv (\phi_i^{(1)} + \phi_i^{(2)} +
\cdots + \phi_i^{(L)} )/ \sqrt{L}$ for $\mid \phi \rangle$ and plot them 
versus $E_k - E$. We see that the values of  
$\langle \phi^{(k)} \mid e^{-\beta (\hat{H}-E)^2} \mid \phi \rangle$
shape a sharp peak in the narrow range around $E_k = E$, while they are 
negligible for other $k$'s.  This powerful selectivity should be emphasized 
to be one of the most advantageous features of our method. 

Finally, we show the results on $ I(E) $ and $\sqrt{\beta/\pi} 
\cdot I(E) $ in Fig.~6, where
the diamonds are the numerical data and  the dashed (solid) line is 
the analytic result with (without) the extra factor 
\footnote{Remember that, 
in order to approximate the delta function by the Gaussian function,
we need the extra factor $\sqrt{\beta/\pi}$.}. 
In the range $200 \le \beta \le 30000$, the Gaussian function multiplied 
by the extra factor is almost flat with the value $\sim 0.5$, 
which is the exact value in the continuum limit $L \rightarrow \infty$.
This result endorses that $\sqrt{\beta/\pi}\cdot I(E) $ fills the role of 
the delta function.

\section{Three-dimensional problems}

In this section we discuss the problems of three dimensions.
Here the Hamiltonian is 
\begin{eqnarray}
\hat{H}= -\vec{\nabla} ^2 +V(\vec{r}) .
\end{eqnarray}
In calculating $exp [ -\beta \{-\vec{\nabla} ^2 + V(\vec{r}) -E\}^2  ]$
%$e^{-\beta \{-\vec{\nabla} ^2 + V(\vec{r}) -E \}^2}$ 
we simply extend the method developed in the one-dimensional case and
add terms of  
$$
 \frac{d^4}{dx^2 dy^2},  \frac{ d^4}{dy^2 dz^2}, \frac{d^4}{dz^2 dx^2},
$$
for which we use the replacements 
\begin{eqnarray}
 \frac{d^4\phi(x,y,z)}{dx^2 dy^2}&\rightarrow&
\frac{1}{\Delta^4} [ \phi_{i+1,j+1,k}+\phi_{i+1,j-1,k}+
 \phi_{i-1,j+1,k}+\phi_{i-1,j-1,k} \nonumber \\
 &-& 2( \phi_{i+1,j,k}+\phi_{i-1,j,k}+ \phi_{i,j+1,k}+\phi_{i,j-1,k})
  +4\phi_{i,j,k} ],
\end{eqnarray}
and so on.

Let us consider the three-dimensional harmonic oscillator, whose 
potential is given by
\begin{eqnarray}
 V(\vec{r}) = \lambda \mid \vec{r} \mid ^2 = \lambda(x^2+y^2+z^2), 
\end{eqnarray}
where we also fix $\lambda = 1/4$ in our calculations.
The energy eigenvalues and their eigenstates can be obtained from
those in the one-dimensional case.
The state $\mid \phi \rangle$ is given by the wavefunction 
$$  \phi_{ijk} = \sum_{i,j,k=1}^{L}
%  = \sum_{1\leq l,m,n\leq L} 
\phi_i^{(l)} \phi_j^{(m)} 
 \phi_k^{(n)} C_{lmn}, $$
where $ \phi_i^{(l)}$ represents an eigenstate in the one-dimensional case
and we take
$$ C_{lmn} = 1/\sqrt{1000}
\ \ {\rm for} \ \ l \leq 10, \ \  m \leq 10,\ \  n \leq 10 . 
$$
Fig.~7 plots the numerical results on
$I(E) = \langle \phi \mid e^{-\beta (\hat{H}-E)^2} \mid \phi \rangle $ 
obtained by the second-order and the fourth-order
decompositions,
which indicates that the error is $O(1/N_t^K)$ in these cases, too. 
We then apply our method to calculate the density of states defined by 
$$   \rho(E) = tr[ \delta(\hat{H} -E) ]. $$
%for the three-dimensional harmonic oscillator.
Replacing the delta function with the Gaussian function 
$$\rho(E,\beta)=tr[ \  \sqrt{\frac{\beta}{\pi}}e^{-\beta(\hat{H} - E)^2} ],
$$ 
we evaluate the $trace$ by means of the random states\cite{iitaka},
$$ \mid \phi \rangle = \sum_{i,j,k=1}^{L} \mid i,j,k \rangle 
\xi_{ijk} , \ \ \ \ 
\langle \! \langle \xi_{ijk}\xi_{i'j'k'} \rangle \! \rangle 
=\delta_{ii'}\delta_{jj'}\delta_{kk'} ,$$
where a state $  \mid i,j,k \rangle $ denotes a particle at the position
$(x_i,y_j,z_k)$ and $ \langle \! \langle \cdot  \rangle \! \rangle  $
represents the statistical average.
Results on $ \rho(E,\beta) $ with our method are shown in Fig.~8 by the dots.
The solid line in the figure gives the exact value of 
$ \rho(E,\beta) $ calculated using 
the all eigenvalues in the discretized three-dimensional space,
$$ \rho(E,\beta) \mid _{exact} =  \sum_{i,j,k=1}^{L}
%\sum_{i=1}^L \sum_{j=1}^L \sum_{k=1}^L
\sqrt{\frac{\beta}{\pi}} e^{  -\beta \{E(i,j,k) - E \}^2  },$$
where $E(i,j,k) = E_i + E_j +E_k$ and $E_{i(j,k)}$ is the $i(j,k)$th energy 
eigenvalue for the one-dimensional harmonic oscillator.  
We also show $E^2$ by the dashed line, 
which is obtained in the continuous space by 
$$ \lim_{\Delta \rightarrow 0} 
\sum _{k_x, k_y, k_z =1} 
\delta ( E - \Delta (\omega_x + \omega_y + \omega_z))
= \int_0^E d\omega'_x d\omega'_y d\omega'_z 
\delta (E- \omega'_x - \omega'_y -\omega'_z),$$
where $\omega_{x(y,z)}$ denotes $k_{x(y,z)} - 1/2$.
We see that the agreement is satisfactory.

\section{Summary}

In this paper we present detailed descriptions of a new method to calculate 
expectation values of a delta function of the Hamiltonian. 
We replace the delta function by the Gaussian function and 
apply the Suzuki-Trotter decomposition to it.
For concrete examples we carry out numerical 
calculations on the quantum mechanical problems.
Our results for the harmonic oscillator
problems in one- and three-dimensional space indicate that this method is 
useful to study the dynamical quantities.

Let us close our summary with a few remarks on further possible developments
of our method. It would be very important to apply
it to the problems beyond the quantum mechanics, the spin system\cite{spin} for
instance.
Applications to the Green function and a step function are
also quite interesting.

\vskip 0.3in
%\eject
\large{{\bf Acknowledgement }}

We would like to thank Profs. M.~Suzuki, S.~Miyashita, H.~Kono, H.~Tanaka and
T.~Nakayama for valuable discussions.
Communications with Dr. Iitaka have been quite helpful to our study.

%\vskip 0.3in
\eject

\vfill
\eject
%\vskip 0.3in

\begin{figure}
\centering
%buna version
%%\epsfile{file=fig1.eps,height=12.0cm,width=14cm}
%\epsfile{file=fig1.eps,scale=0.45}
%kiri version
\epsfxsize=0.5\textwidth
\epsfbox{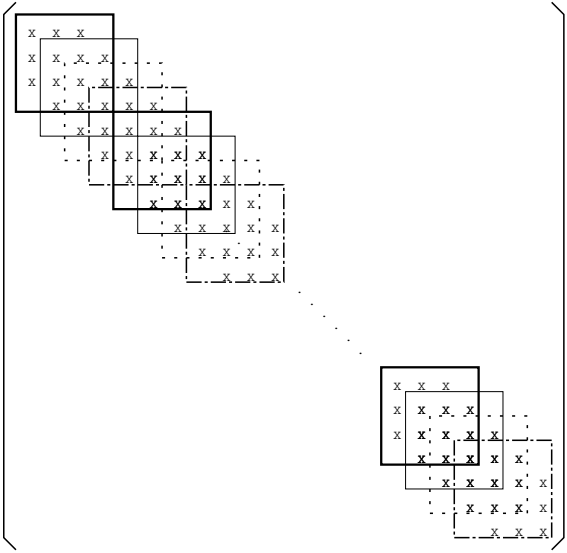}
\caption{
A schematic picture for the matrix $H_2$, where crosses (x) indicate its 
non-zero matrix elements. 
$H_2^{(1)}$ ($H_2^{(2)}$,  $H_2^{(3)}$, $H_2^{(4)}$) has 
non-zero elements only within the area surrounded by the bold (solid, 
dotted, dot-dashed) line.
}
\end{figure}

\begin{figure}
\centering
%\epsfile{file=fig2.eps,scale=0.45}
\epsfxsize=0.65\textwidth
\epsfbox{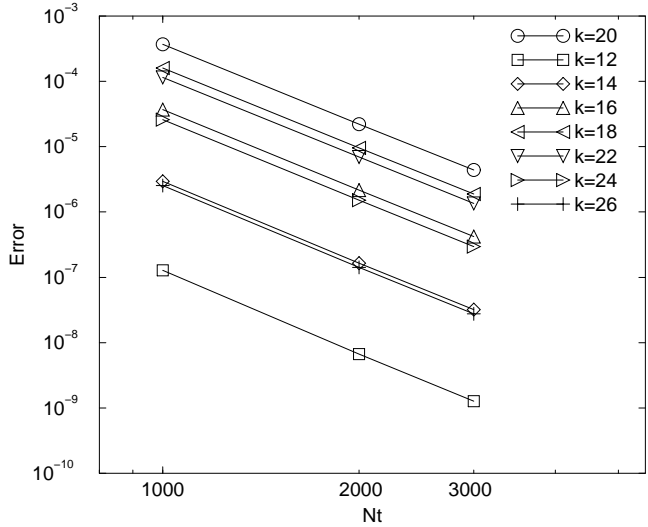}
\caption{
Errors in our calculations on 
$ \langle \phi^{(k)} \mid e^{ -\beta(\hat{H}-E)^2 } \mid \phi^{(20)} 
\rangle$ in the fourth-order
Suzuki-Trotter decomposition with $\beta =100$ and $E=0.094716$ 
versus the Trotter number $N_t$. Here
$\mid \phi^{(k)} \rangle$ denotes the $k$th eigenstate of the Hamiltonian
and the error is defined by the absolute value of
$\langle \phi^{(k)} \mid e^{ -\beta(\hat{H}-E)^2 } \mid \phi^{(20)} 
\rangle - e^{-\beta(E_{20}-E)^2} \cdot \delta_{k,20} $ for each $k$. 
}
\end{figure}

\begin{figure}
\centering
%\epsfile{file=fig3.eps,scale=0.45}
\epsfxsize=0.65\textwidth
\epsfbox{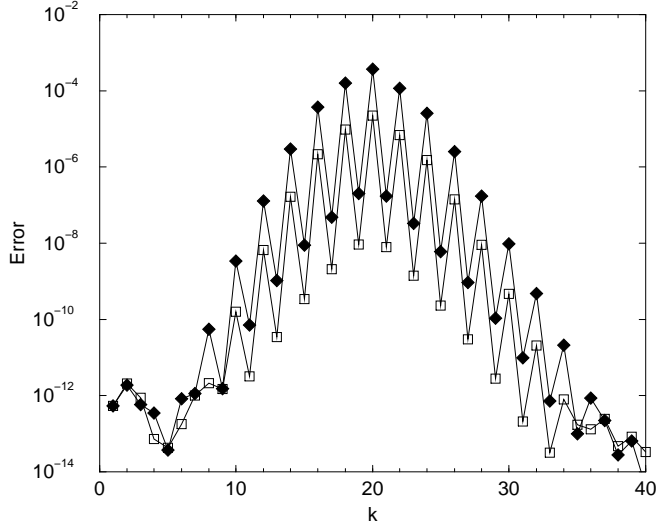}
\caption{
 Errors on 
$ \langle \phi^{(k)} \mid e^{ -\beta(\hat{H}-E)^2 } \mid \phi^{(20)} 
\rangle$
in the fourth-order decomposition for $k \le 40$  
with fixed Trotter numbers $N_t= 1000$ (diamonds) and 
$N_t = 2000$ (squares). Values of the parameters 
$\beta$ and $E$ are the same as those employed in Fig.~2.  
}
\end{figure}

\begin{figure}
\centering
%\epsfile{file=fig4.eps,scale=0.45}
\epsfxsize=0.65\textwidth
\epsfbox{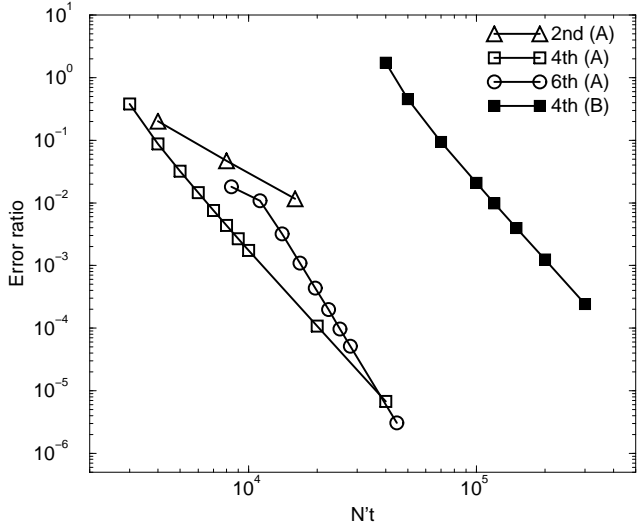}
\caption{ The error ratio of $I(E)$ with $E = 0.094716$ defined by
(the approximate value $-$ the exact value)/(the exact value)
with the second-, the fourth- and 
the sixth-order formulae versus the effective Trotter number $N'_t$. 
The value of $\beta$ is $200 $ for the wavefunction (A),
while it is $2000$ for the wavefunction (B).
}
\end{figure}

\begin{figure}
\centering
%\epsfile{file=fig5.eps,scale=0.45}
\epsfxsize=0.65\textwidth
\epsfbox{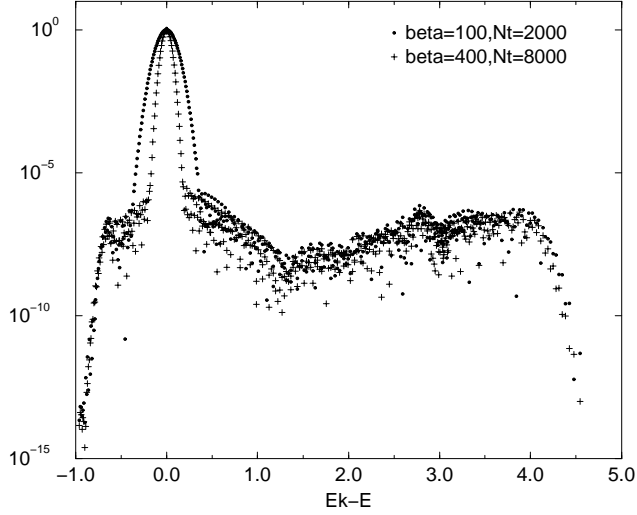}
\caption{Absolute values of 
$ L \cdot \langle \phi^{(k)} \mid e^{-\beta (\hat{H}-E)^2} \mid \phi \rangle$
($1 \le k \le L$ and $E = 0.96302$) plotted as a function of $E_k - E$ 
for $\beta = 100$, $N_t = 2000$ (the dots) and 
for $\beta = 400$ and $N_t=8000$ (the pluses),
using the fourth-order decomposition.
}
\end{figure}

\begin{figure}
\centering
%\epsfile{file=fig6.eps,scale=0.45}
\epsfxsize=0.65\textwidth
\epsfbox{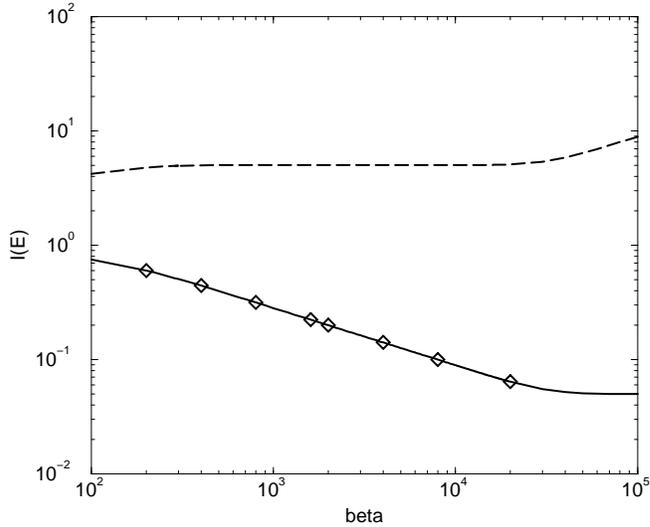}
\caption{The numerical data (the diamonds) and the analytic result (the
solid line) on $I(E)$. The analytic result on
% $\sqrt{\beta/\pi} $ .
 $(\beta/\pi)^{1/2} \cdot I(E)$ are also presented by the dashed line.
Here we use the wavefunction (B) with $E=0.094716$.
The Trotter number $N_t$ is fixed to be  $20\beta$.
}
\end{figure}

\begin{figure}
\centering
%\epsfile{file=fig7.eps,scale=0.45}
\epsfxsize=0.65\textwidth
\epsfbox{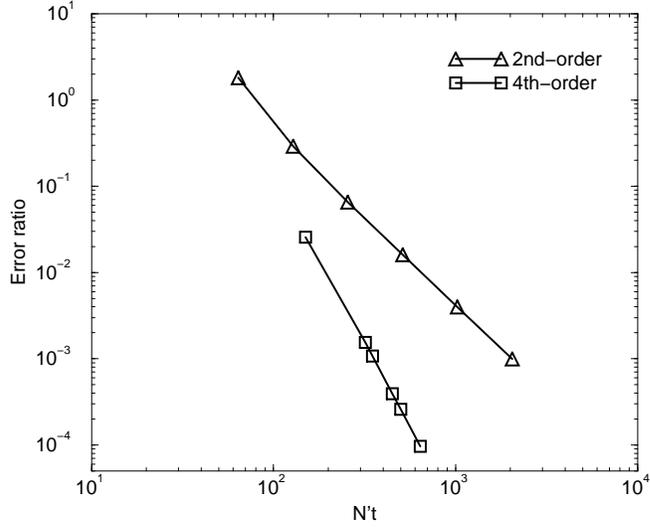}
\caption{ The error ratio of $I(E)$ versus the effective Trotter number
 $N'_t$ for the three-dimensional harmonic oscillator.
  Here $L^3 = 32^3 $,  $x_{max}=-x_{min}=8.0$ ,
  $\beta  =  4.0$ and     $  E=   1.1101$.
%The number $N'_t$ is the Trotter number $N_t$ itself for the second-order 
%decomposition, while it is $5 N_t$ for the fourth-order one.
}
\end{figure}

\begin{figure}
\centering
%\epsfile{file=fig8.eps,scale=0.45}
\epsfxsize=0.65\textwidth
\epsfbox{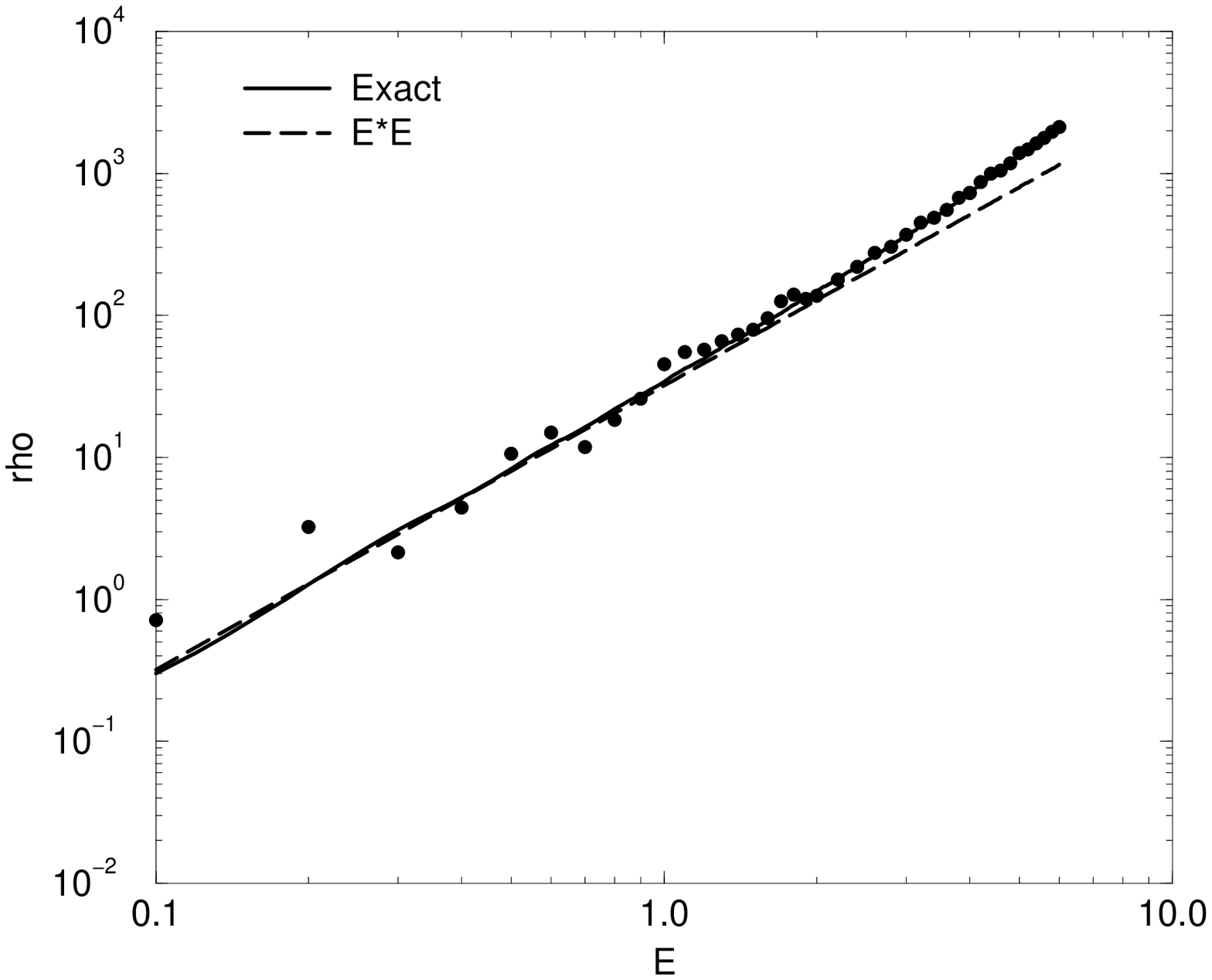}
\caption{  $\rho(E,\beta)$ with  $\beta  =  32.0$ versus the energy $E$ for
  the three-dimensional
harmonic oscillator.
  Here $L^3 = 32^3 $,  $x_{max}=-x_{min}=8.0$ and $  N_t=300$.
Results of our calculation (dots) are obtained by the averages over
10 samples of the random states.
}
\end{figure}


\begin{thebibliography}{99}

\bibitem{iitaka}

 T. Iitaka: RIKEN Review {\bf 19} (1997) 136.
   
\bibitem{rv}
T. Nakayama and H. Shima: Phys. Rev. E{\bf 58} (1998) 3984;
J. Phys. Soc. Jpn. 67 (1998) 2189.

T. Iitaka et al.: Phys. Rev. E{\bf 56} (1997) 1222,

T. Iitaka: Phys. Rev. E{\bf 56} (1997) 7318,

H. Tanaka: Phys. Rev. B{\bf 57} (1998) 2168.

T. Roche: cond-mat/9805369.

M. Itoh, Phys. Rev. B{\bf 45} (1992) 4242.

\bibitem{delta}
R. N. Silver, H. Roeder, A. F. Voter and J. D. Kress:
J. Compt. Phys. {\bf 124} (1996) 115.

R .Chen and H. Guo: Chem. Phys. Lett. {\bf 261} (1996) 605;
Compt. Phys. Commun. {\bf 119} (1999) 19.

\bibitem{filter}

H. Kono: Chem. Phys. Lett. {\bf 214} (1993) 137.

R. E. Wyatt: Phys. Rev. E.{\bf 51} (1995) 3643.

C. Iung and C. Leforestier: J. Chem. Phys. {\bf 102} (1995) 8453.

M. R. Wall and D. Neuhauser: J. Chem. Phys. {\bf 102} (1995) 8011.

\bibitem{MC}
 N. Hatano and M. Suzuki: in 
{\it Quantum Monte Carlo Methods in Condensed Matter},
edited by M.\ Suzuki, (World Scientific, Singapore, 1993), p.13.

\bibitem{natori}
 H. Natori and T. Munehisa: J. Phys. Soc. Jpn. {\bf 66} (1997) 351. 

H. De Raedt and K. Michielsen: Comp. Phys. {\bf 8} (1994) 600.

\bibitem{suzuki}

 M. Suzuki: Commun. Math. Phys. {\bf 51} (1976) 183;
 in {\it Quantum Monte Carlo Methods in Condensed Matter},
edited by M.\ Suzuki, (World Scientific, Singapore, 1993), p.1.

\bibitem{spin}
S. Miyashita, T. Yoshino and A. Ogasahara: J. Phys. Soc. Jpn. {\bf 68}
 (1999) 655.
\end{thebibliography}
\end{document}